\documentclass[12pt,preprint]{aastex}

\slugcomment{Draft version: Feb. 26, 2012 (revision)}

\shorttitle{Suzaku Observation on Kookaburra region}
\shortauthors{Kishishita et al.}

\begin{document}


\title{X-ray investigation of the diffuse emission around plausible $\gamma$-ray emitting pulsar wind nebulae in Kookaburra region}


\author{Tetsuichi Kishishita\altaffilmark{1}, Aya Bamba\altaffilmark{2}, Yasunobu Uchiyama\altaffilmark{3, 4}, Yasuyuki Tanaka\altaffilmark{5}, and Tadayuki Takahashi\altaffilmark{5, 6}}


\altaffiltext{1}{Universit$\ddot{{\rm a}}$t Bonn, Physikalisches Institut, Nussallee 12, Bonn 53115, Germany}
\altaffiltext{2}{Department of Physics and Mathematics, College of Science and Engineering, Aoyama Gakuin University, 5-10-1, Fuchinobe, Chuo-ku, Sagamihara, Kanagawa, 252-5258, Japan}
\altaffiltext{3}{Kavli Institute for Particle Astrophysics and Cosmology, SLAC National Accelerator Laboratory
2575 Sand Hill Road M/S 29, Menlo Park, CA 94025}
\altaffiltext{4}{Panofsky Fellow}
\altaffiltext{5}{Institute of Space and Astronautical Science, Japan Aerospace Exploration Agency, Sagamihara, Kanagawa 229-8510, Japan}
\altaffiltext{6}{Department of Physics, The University of Tokyo, Bunkyo, Tokyo 113-0033, Japan}


\begin{abstract}
We report on the results from {\it Suzaku} X-ray observations of the radio complex region called Kookaburra, which includes two adjacent TeV $\gamma$-ray sources HESS~J1418-609 and HESS~J1420-607.  The {\it Suzaku} observation revealed X-ray diffuse emission around a middle-aged pulsar PSR~J1420-6048 and a plausible PWN Rabbit with elongated sizes of $\sigma_{\rm X}=1^{\prime}.66$ and $\sigma_{\rm X}=1^{\prime}.49$, respectively. The peaks of the diffuse X-ray emission are located within the $\gamma$-ray excess maps obtained by H.E.S.S. and the offsets from the $\gamma$-ray peaks are $2^{\prime}.8$ for PSR~J1420-6048 and $4^{\prime}.5$ for Rabbit. The X-ray spectra of the two sources were well reproduced by absorbed power-law models with $\Gamma=1.7-2.3$. The spectral shapes tend to become softer according to the distance from the X-ray peaks. Assuming the one zone electron emission model as the first order approximation, the ambient magnetic field strengths of HESS~J1420-607 and HESS~J1418-609 can be estimated as 3~$\mu$G, and $2.5~\mu$G, respectively. 
The X-ray spectral and spatial properties strongly support that both TeV sources are pulsar wind nebulae, 
in which electrons and positrons accelerated at termination shocks 
of the pulsar winds are losing 
 their energies via the synchrotron radiation and inverse Compton scattering 
 as they are transported outward.
\end{abstract}





\section{INTRODUCTION}
Gamma-ray emitting Pulsar Wind Nebulae (PWNe) are excellent test grounds for studying pulsars' relativistic winds and particle acceleration that takes place presumably at termination 
shocks. 
Especially, PWNe which accompany the diffuse synchrotron X-ray emission show the important clues to understand the time evolution of the PWNe and how the accelerated particles escape from the shocks. 
While young PWNe with characteristic ages of less than 10~kyr are relatively well studied in the past observations with {\it Chandra} and {\it XMM}-{\it Newton} \citep{kargaltsev, bambaa}, the sample of middle aged PWNe remains small. The Kookaburra region includes a middle aged PWN (PSR~J1420-6048) and a plausible PWN (Rabbit), making it a suitable target for the detailed study of the diffuse emission with the {\it Suzaku} X-ray observatory.
{\it Suzaku}, characterized by the low detector background compared to {\it Chandra} and {\it XMM-Newton}, is crucial for the analysis of faint and diffuse X-ray emission. 

The complex of compact and extended radio/X-ray sources, called Kookaburra (designated by \citealt{robert99}), spans over about one square degree along the Galactic plane around $l=313^{\circ}.4$.  The H.E.S.S. galactic survey revealed two very high energy (VHE) sources in this region; the brighter of the two sources, HESS~J1420-607, is centered at the position of (R.A., Dec.) = (14$^{\rm h}$20$^{\rm m}$09$^{\rm s}$, -60$^{\circ}$45$^{\prime}$36$^{\prime\prime}$) with an intrinsic extension of $\sigma_{\rm TeV}=3^{\prime}.3\pm0^{\prime}.5$, and the slightly less bright second source, HESS~J1418-609, is centered at the position of (R.A., Dec.) = (14$^{\rm h}$18$^{\rm m}$04$^{\rm s}$, -60$^{\circ}$58$^{\prime}$31$^{\prime\prime}$) with an intrinsic extension of $\sigma_{\rm TeV}=3^{\prime}.4\pm0^{\prime}.6$ (a major-axis of $4^{\prime}.9\pm1^{\prime}.5$ and a minor-axis of $2^{\prime}.7\pm0^{\prime}.7$ fitted with an elongated Gaussian shape). 
The Kookaburra region is also bright in the GeV band and registered as a {\it Fermi} $\gamma$-ray source \citep{welt}.

There is an energetic pulsar PSR~J1420-6048 at the south of HESS~J1420-607. It is located at (R.A., Dec.) = (14$^{\rm h}$20$^{\rm m}$08$^{\rm s}.20$, -60$^{\circ}$48$^{\prime}$17$^{\prime\prime}.2$), which is $\sim3^{\prime}.1$ offset from the center of the VHE $\gamma$-ray emission. PSR~J1420-6048 is a radio/X-ray pulsar with a period of $P=68$~ms and period derivative of $\dot{P}=8.3 \times 10^{-14}$ \citep{damico, ng}. The distance of the pulsar was estimated as $d=5.6$~kpc based on the pulsar's dispersion measure with the NE2001 Galactic electron-density model \citep{cordes}. The characteristic age and the spin-down luminosity are $\tau=13$~kyr and $\dot{E}=1.0 \times 10^{37}$~erg s$^{-1}$, respectively. The strength of surface magnetic field is $B_s = 2.5 \times 10^{12}$ G. 

{\it Chandra} observations revealed a faint and narrow diffuse X-ray emission in K3 nebula (designated by \citealt{robert99} with the radio observation) around PSR~J1420-6048 \citep{ng}. Spectral fitting with an absorbed power-law model to the K3 nebula ($2^{\prime}$ radius from the pulsar) yielded an absorption column density of $N_{\rm H}=5.4^{+2.2}_{-1.7} \times 10^{22}~{\rm cm}^{-2}$ with a photon index of $\Gamma=2.3^{+0.9}_{-0.8}$.

The Rabbit nebula is located at the Eastern edge of HESS~J1418-609 with a distance of $8^{\prime}.2$ to the best central position of the VHE emission. \cite{ng} found two point-like sources with the {\it XMM-Newton} data, labeled as R1 and R2; the brighter source (R1) is located at the edge of the Rabbit, while the fainter source (R2) at (R.A., Dec.) = (14$^{\rm h}$18$^{\rm m}$39$^{\rm s}.90$, -60$^{\circ}$57$^{\prime}$56$^{\prime\prime}.5$), which appears embedded in the narrow diffuse emission. Timing analysis with the EPIC pn data revealed a $P=108$~ms pulsation with a period derivative of $\dot{P}=1.07 \times 10^{-12}$ from R2, although not highly significant. Assuming these values are correct, the characteristic age can be estimated as $\tau=1.6$~kyr. Fitting the spectrum from the whole Rabbit nebula ($3^{\prime}$ radius) with an absorbed power-law model, they found an absorption column density of $N_{\rm H}=1.4 \pm 0.2 \times 10^{22}~{\rm cm}^{-2}$ and a photon index of $\Gamma=1.5 \pm 0.14$.

The two VHE sources are considered to be PWNe \citep{robert99, ng}, given 
the X-ray and TeV results reported so far. 
Generally, detailed observations of the synchrotron X-ray nebulae are crucial to 
understand spatial and spectral distributions of accelerated electrons. 
However, previous X-ray observations did not have enough sensitivity to detect 
faint X-ray nebulae with sizes comparable to the TeV nebulae. 
In this paper, we report new detections of the diffuse X-ray emission of the Kookaburra complex using the {\it Suzaku} satellite.

\section{OBSERVATION AND DATA REDUCTION}
We observed the Kookaburra region with the {\it Suzaku} satellite in 2009 February. 
The {\it Suzaku} observations were performed with the X-ray Imaging Spectrometer (XIS: \citealt{koyama_xis06}) in 0.3 -- 12~keV and the Hard X-ray Detector (HXD: \citealt{takahashi_hxd06}) in 13 -- 600~keV. The XIS, located at the focal plane of the X-ray telescopes (XRT: \citealt{serle07}), consists of one back-illuminated CCD camera (XIS1) and three front-illuminated CCDs (XIS0, 2, and 3). One of the front-illuminated CCDs, XIS2, was not available at the time of our observations, since it suffered from a fatal damage on 9 November 2006, and unusable since then. 
The XIS instruments were operated in a normal full-frame clocking mode (a frame time of 8~s) with Spaced-row Charge Injection (SCI) \citep{nakajima}. 
The HXD consists of the silicon PIN photo diodes (hereafter PIN) capable of observations in the 13 -- 70~keV band and the GSO crystal scintillators (hereafter GSO) which cover the 40 -- 600~keV band. Since we could not find significant signals in the HXD data after subtracting the non-X-ray background (NXB), cosmic X-ray background (CXB), and galactic ridge components, we focused on the XIS data analysis in this paper. 

We used data sets processed by a set of software of the {\it Suzaku} data processing pipeline (version 2.2.11.24). The telemetry saturating time was excluded in the pipeline processing. Basic analysis was done using the HEASOFT software package (version 6.4). 
We made use of cleaned event files, in which standard screening was applied. The standard screening procedures include event grade selections, and removal of time periods such as spacecraft passage of the South Atlantic Anomaly (SAA), intervals of small geomagnetic cutoff rigidity (COR), and those of a low elevation angle. Specifically, for the XIS, the elevation angle larger than 5$^{\circ}$ above the Earth and larger than 20$^{\circ}$ from the sunlit Earth limb are selected. 
The total exposure from the two observations amounts to 72~ks after standard data screening. Table \ref{tab:kookaburra_obs} gives the log of the {\it Suzaku} observations.

\begin{table}[htdp]
\caption{Summary of the {\it Suzaku} observation on the Kookaburra region}
\begin{center}
\begin{tabular}{cccc}
\hline
\hline 
Obs. ID & Coord. (J2000) & Exposure \tablenotemark{a} & Date \\ 
  & R.A., Dec. & XIS & \\
 \hline
503110010 & 215.0292, -60.8156 & 50.3 & 01/11/2009\\
503071010 & 214.6625, -60.9675 & 21.3 & 02/14/2009\\
\hline
\tablenotetext{a}{Units are ks.}
\end{tabular}
\end{center}
\label{tab:kookaburra_obs}
\end{table}%

\section{ANALYSIS AND RESULTS}

\subsection{Image Analysis}
\begin{figure}
\epsscale{1.0}
\plotone{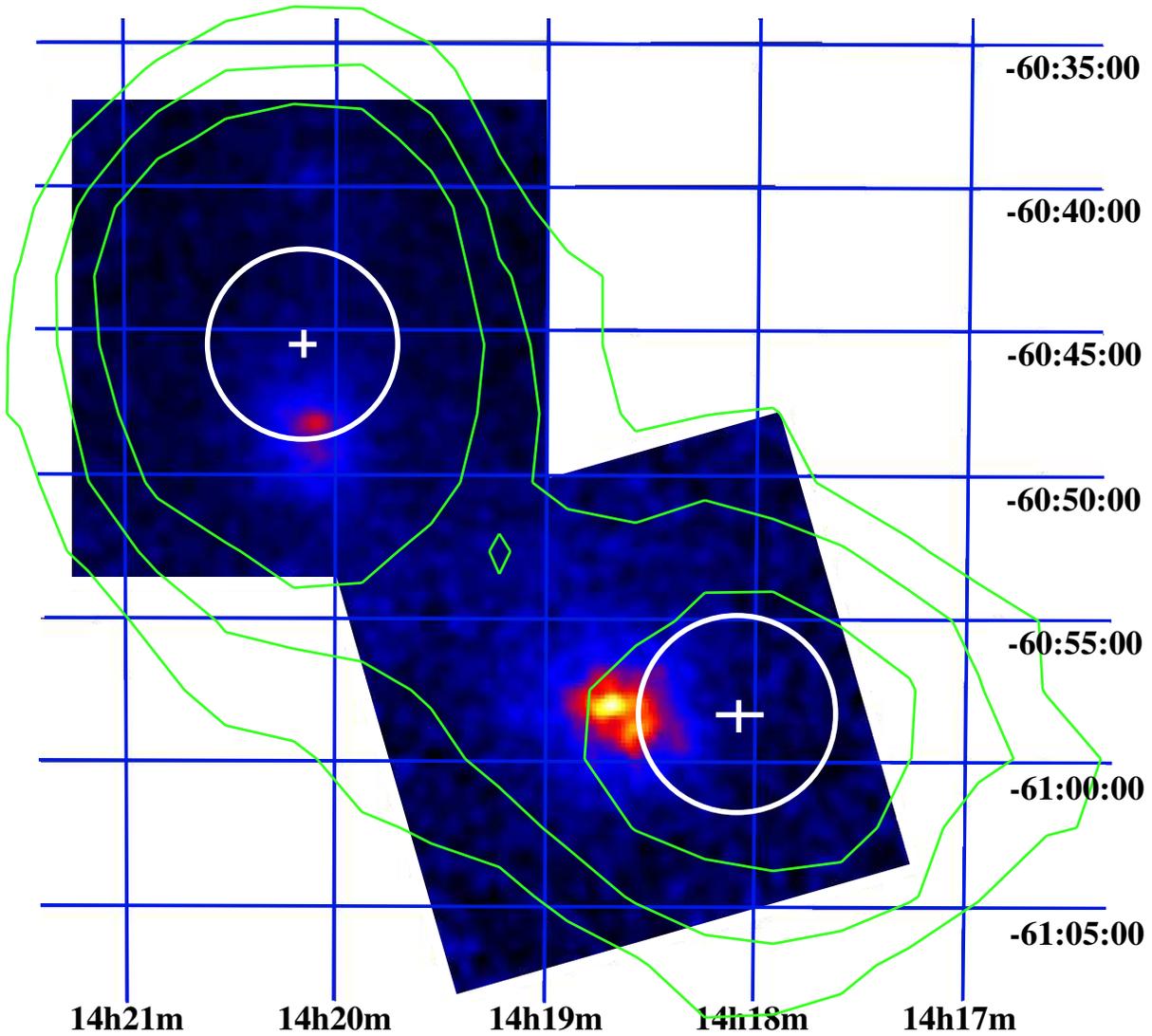}
\caption{{\it Suzaku} XIS (0+1+3) image around the Kookaburra region in a linear scale. The extracted energy range is 2--10~keV. Green contours denote the 5~$\sigma$, 7.5~$\sigma$ and 10~$\sigma$ significance levels reported by H.E.S.S. \citep{aha06}. The white indicates the best-fit positions and intrinsic extensions of the TeV-$\gamma$-rays.}
\label{fig:image_merge}
\end{figure}

In order to combine two different pointing images, we extracted the photon events in the energy range of 2--10~keV from each sensor. The data between 5.73 and 6.67~keV were removed from the image to exclude the calibration sources. We corrected the vignetting effect by dividing the image with a flat sky image simulated in the energy range of 2--8~keV using the XRT + XIS simulator \citep{ishisaki}. The image was binned to 8 $\times$ 8 pixels and smoothed with a Gaussian function of $3\sigma$. A combined {\it Suzaku} XIS (0+1+3) image in the Kookaburra region is show in Fig. \ref{fig:image_merge}. {\bf Both HESS~J1420-607 and HESS~J1418-609 are reported as diffuse TeV-$\gamma$-ray emitting sources \citep{aha06}. The intrinsic extensions of the diffuse emissions are indicated as white circles in the figure.
The overlaid contours are taken from the H.E.S.S. observation.
We can clearly see that the separate X-ray sources locate with some offsets from the $\gamma$-ray peak positions and are coincident with the VHE sources.}

\subsubsection*{K3/PSR~J1420-6048}
\begin{figure*}
\begin{center}
  \includegraphics[width=0.7\hsize,clip]{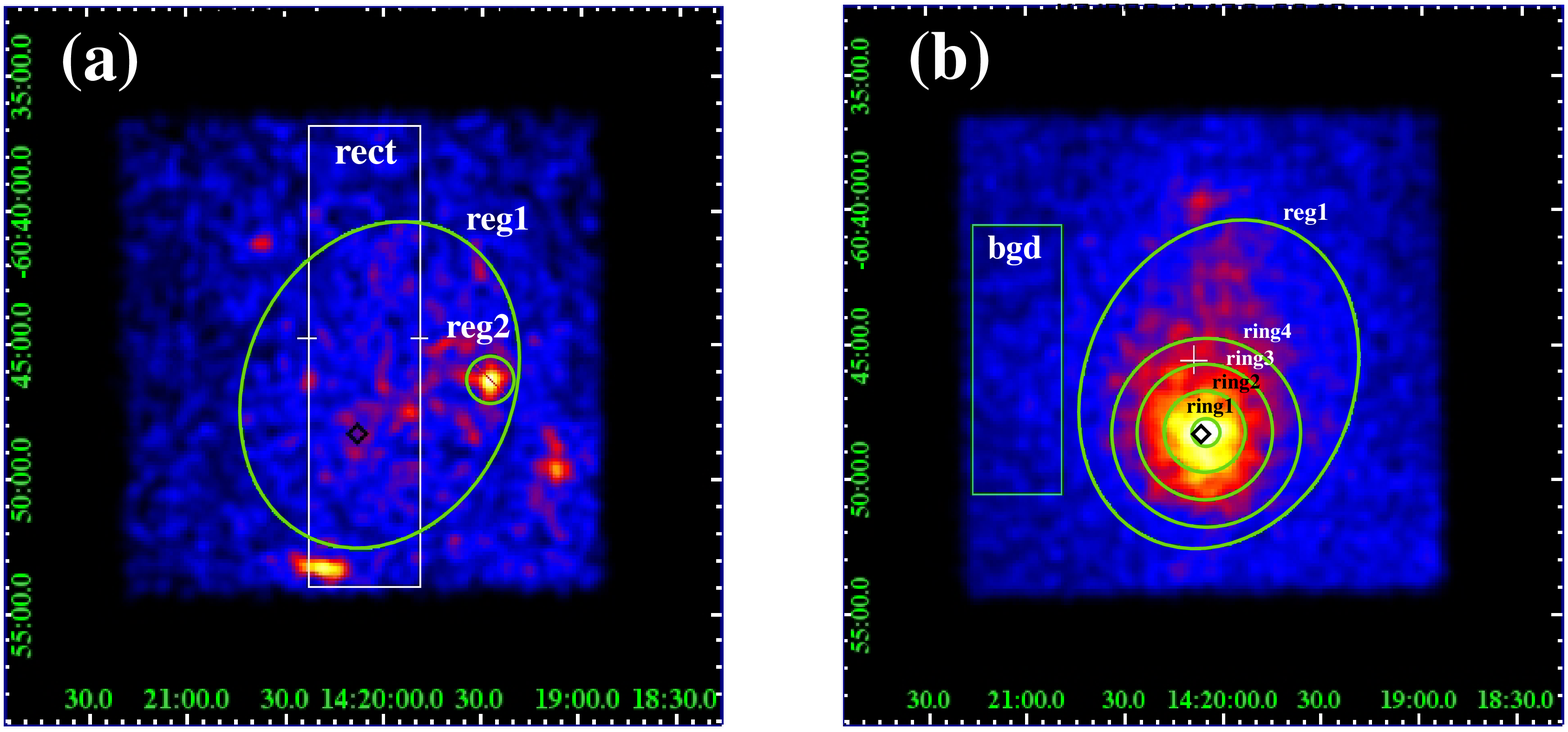}
\caption{XIS images around K3/PSR~J1420-6048 in the energy range of (a) 1--2~keV and (b) 2--10~keV. PSR~J1420-6048 is depicted as a black diamond. The white cross point corresponds to the best-fit position of VHE $\gamma$-rays \citep{aha06}.}
\label{fig:K3image}
\end{center}
\end{figure*}

\begin{figure}
 \includegraphics[width=0.9\hsize,clip]{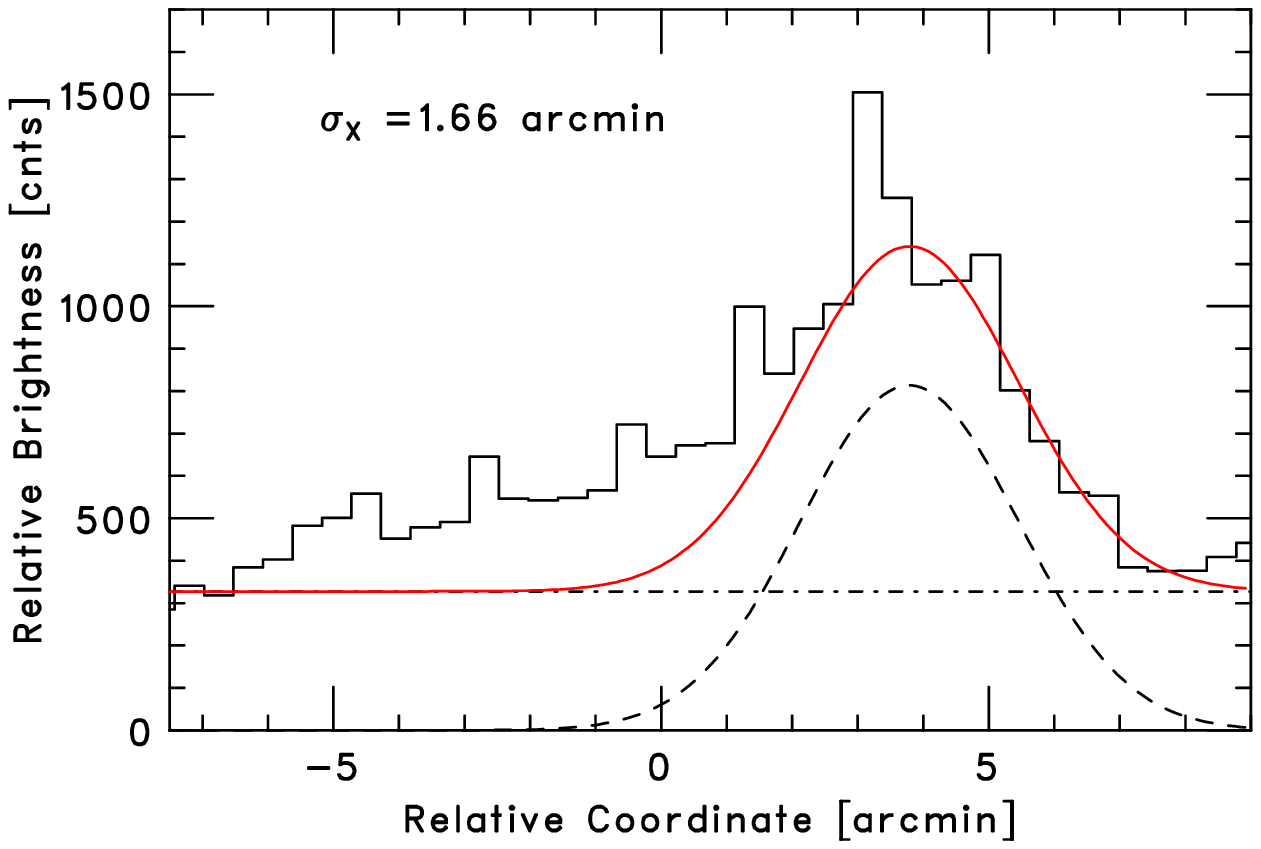}
\caption{Projected profile of the surface brightness around K3/PSR~J1420-6048. The extracted region is shown as the white rectangle ("rect") in Fig. \ref{fig:K3image} (a).}
\label{fig:1D_K3pos}
\end{figure}

We extracted the photon events in the energy ranges of 1--2~keV and 2--10~keV.
The XIS images are shown in Fig. \ref{fig:K3image}. 
We can see several thermal sources in the lower energy image, which are invisible in the higher energies. 
The bright X-ray emission is well coincident with the location of PSR~J1420-6048 and extended diffuse emission surrounds the bright pulsar.
The diffuse component has an elongated shape which extends from the pulsar to the $\gamma$-ray peak.
{\bf In order to determine the extension of non-thermal X-ray emission,
we created a surface brightness profile from the enclosed region shown as "rect" in Fig. \ref{fig:K3image} (a) along the north to south direction. The surface brightness in 2--10~keV vs. relative coordinate is shown in Fig. \ref{fig:1D_K3pos}. The profile was fitted with a Gaussian function plus constant as a background to derive the size of the diffuse emission, using the sigma of the Gaussian ($\sigma_{\rm X}$). 
Since a Gaussian profile often provides a reasonable approximation for the surface brightness profiles \citep{bambab}, we used relative coordinates between 2$^{\prime}$.2 and 9$^{\prime}$.0 as a fitting range in consideration for the non-symmetrical profile. We excluded a range between 2$^{\prime}$.5 and 4$^{\prime}.0$ as a narrow pulsar component. This is to avoid the contribution from the bright pulsar, whose image is smeared by the point-spread function of XRT.
We defined the source size as three times the $\sigma_{\rm X}$. The fitted result shows $\sigma_{\rm X}=1^{\prime}.66\pm0^{\prime}.09$. If we assume the distance as 5.6~kpc, the physical size of the extended diffuse emission corresponds to $8.1\pm0.4$~pc.
The intrinsic emission size of the bright pulsar was also estimated by fitting the image with a two-dimensional Gaussian function, however, the calculated size was smaller than the attitude fluctuation of the satellite \cite[see][]{attitude}.}

\subsubsection*{Rabbit}
\begin{figure*}
  \begin{center}
 \includegraphics[width=0.7\hsize,clip]{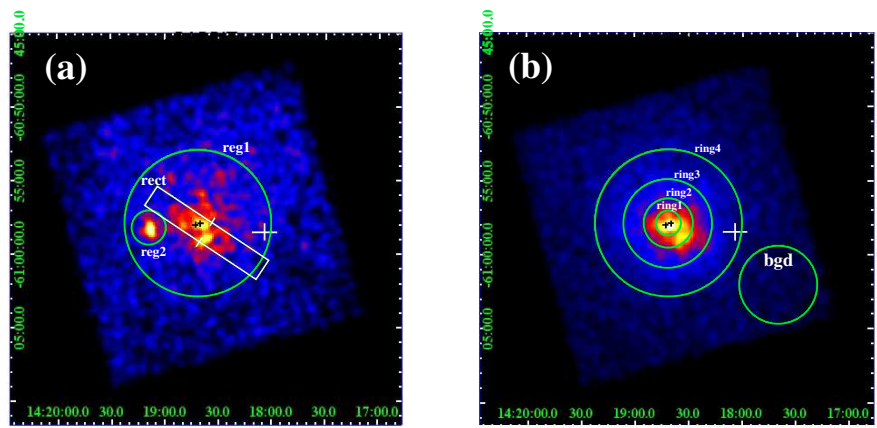}
\caption{XIS images around the Rabbit nebula in the energy range of (a) 1--2~keV and (b) 2--10~keV. The black cross points show the point-like sources detected by {\it XMM}-{\it Newton}, denoted as "R1" and "R2" in \cite{ng}. The white cross point corresponds to the best-fit position of VHE $\gamma$-rays \citep{aha06}.}
\label{fig:Raimage}
\end{center}
\end{figure*}

\begin{figure}
 \includegraphics[width=0.9\hsize,clip]{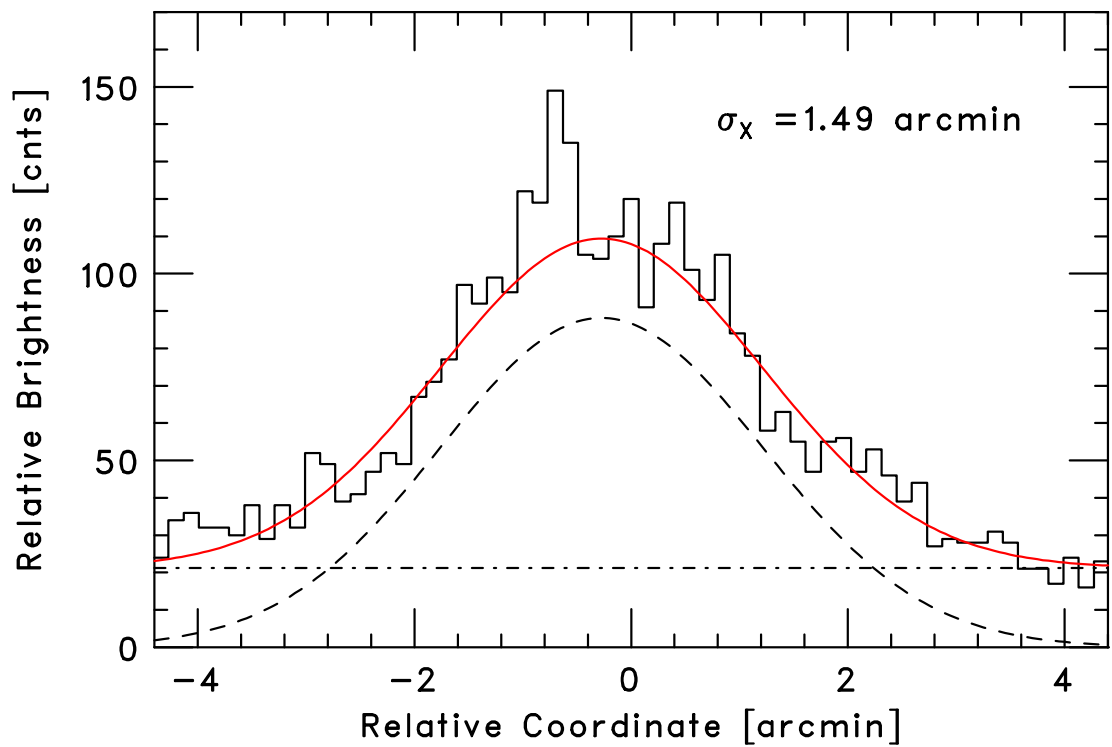}
\caption{Projected profile of the surface brightness around the Rabbit nebula. The extracted region is shown as the white rectangle ("rect") in Fig. \ref{fig:Raimage} (a).}
\label{fig:1D_Rapos}
\end{figure}

Fig. \ref{fig:Raimage} shows the XIS images of the Rabbit nebula.
Similar to K3/PSR~J1420-6048, thermal sources are visible in the lower energies. 
The X-ray peak is coincident with the point-like sources reported by \cite{ng}, which are indicated as black cross points in Fig. \ref{fig:Raimage}. According to \cite{ng}, these bright sources are embedded in narrow diffuse emission. Compared with the {\it XMM}-{\it Newton}, the diffuse emission in Fig. \ref{fig:Raimage} (b) looks more extended than the previous image, and another diffuse structure which is clearly separated from the point-like sources can be seen in the southeast direction. In order to determine the size of the diffuse emission, we made a projected profile with the same procedure as in K3/PSR~J1420-6048. The extracted area is indicated as "rect" in Fig. \ref{fig:Raimage} (a). The 1D projected result is shown in Fig. \ref{fig:1D_Rapos}. We excluded a range between $-1^\prime .2$ and $-0^{\prime}.4$ from the fitting as the central pulsar region to avoid its influence.
The projected profile was well represented with a Gaussian function plus constant background component. The size of the diffuse component is estimated as $\sigma_{\rm X}=1^{\prime}.49\pm0^{\prime}.07$ from the 1D profile. The physical size of the diffuse emission can be estimated as $6.5\pm0.3$~pc.
We assumed the distance as 5~kpc.

\subsection{{\bf Spectral Analysis}}

\subsubsection*{K3/PSR~J1420-6048}
Firstly, we checked the background level extracted from the "bgd" region in Fig. \ref{fig:K3image} (b). We compared its background spectrum with public blank sky observations on Lockman-hole (Obs. ID: 102018010) and also with ASO~0304 (Obs. ID: 504054010), which locates near the Kookaburra region, i.e. (R.A., Dec.)=(213.3355, -62.0808). The background levels are consistent within 10\% and the Galactic ridge emission is ignorable. We thus concluded that the "bgd" region is a source-free in the following spectral analysis.

\begin{figure}
 \includegraphics[width=0.9\hsize,clip]{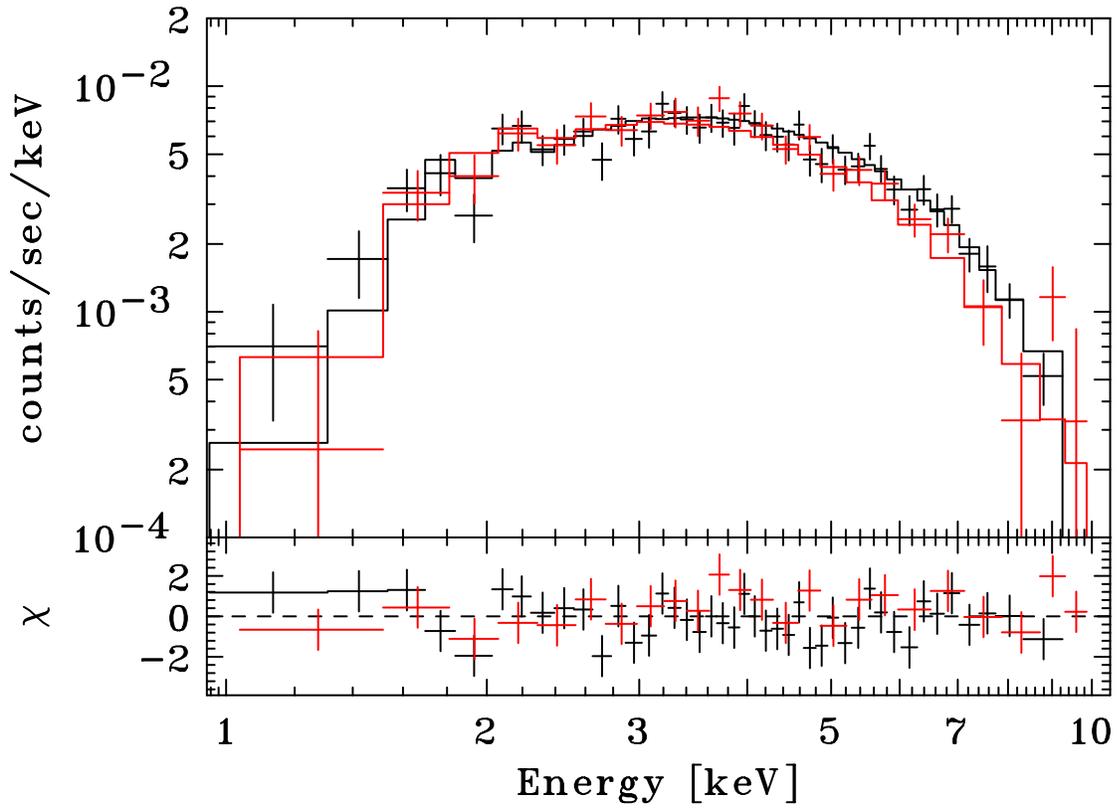}
\caption{Spectra of the "ring1+ring2" in K3/PSR~J1420-6048 (black: XIS0+3, red: XIS1). Solid lines show the fitted absorbed power-law model. The bottom panels show residuals from the best-fit.}
\label{fig:spec_K3_src}
\end{figure}

In order to determine the absorption column density around PSR~J1420-6048, we extracted the photon events within an angular distance of 1$^{\prime}.5$ from 
 the pulsar, which corresponds to "ring1+ring2" in Fig. \ref{fig:K3image} (b). 
We have co-added the data from XIS0 and XIS3 to increase statistics.
The response (rmf) files and the auxiliary response (arf) files were produced using {\it xisrmfgen} and {\it xissimarfgen}, respectively.
Fig. \ref{fig:spec_K3_src} shows the spectral result with the joint-fitted XIS0+3 (black) and XIS1 (red) data. The spectral shapes are well represented by an absorbed power-law model (`wabs$\times$pow'). From the spectral fitting, the absorption column density was estimated as ${\rm N_H}=4.4\pm0.3\times10^{22}~{\rm cm}^{-2}$, which is consistent with the previous observations.
To determine the total flux from the pulsar and diffuse emission, we used the "reg1-reg2" as the extracted region. The flux in the energy range of 2--10~keV is $4.56\pm0.10\times10^{-12}$~erg$\cdot{\rm cm}^{-2}\cdot{\rm s}^{-1}$. The systematic uncertainty coming from the difference of the effective area between the arf responses for diffuse and point sources was less than 10\%.

\begin{figure}
 \includegraphics[width=0.9\hsize,clip]{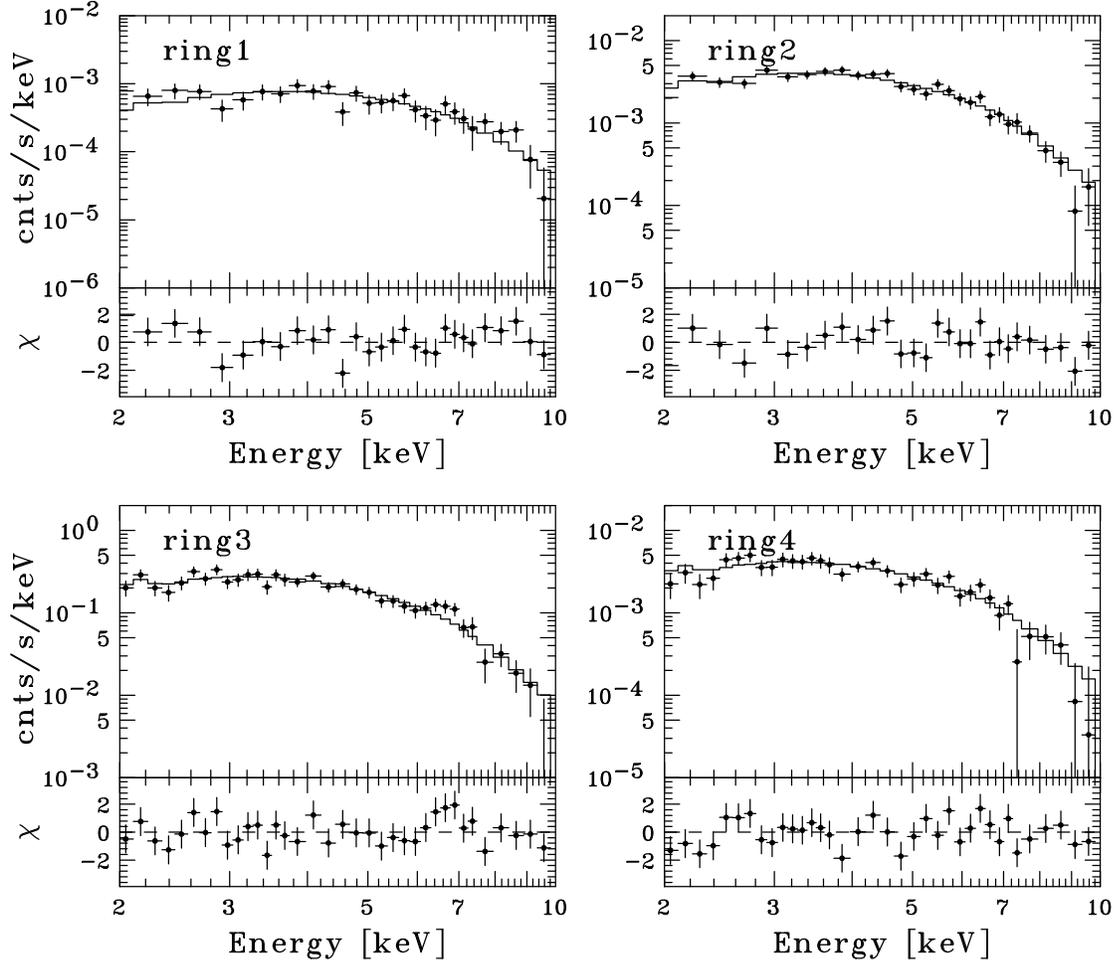}
\caption{Spectra of the circular regions in K3/PSR~J1420-6048. The extracted regions are shown in Fig. \ref{fig:K3image} (b).}
\label{fig:K3_4specs}
\end{figure}

\begin{figure}
  \includegraphics[width=0.9\hsize,clip]{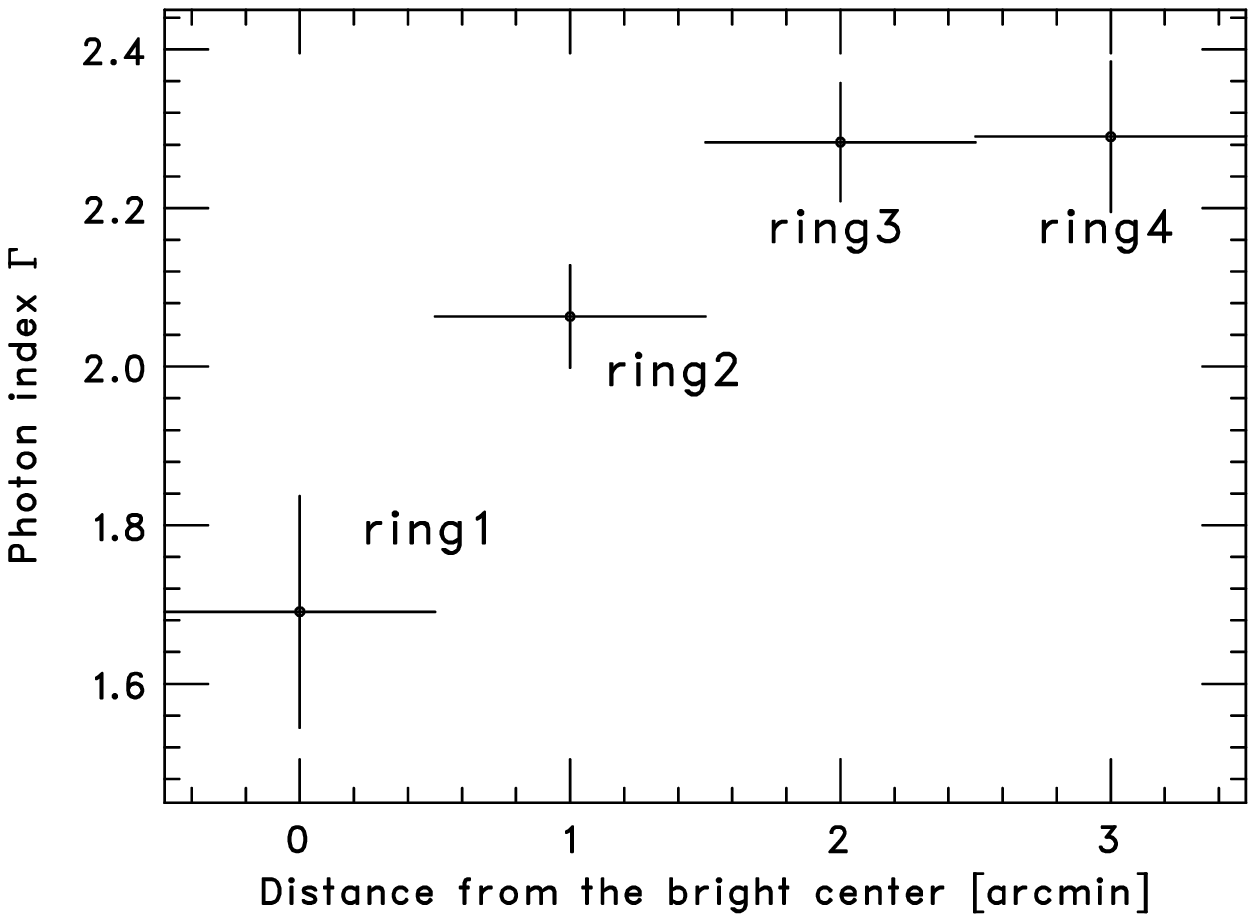}
\caption{Spatial dependence of the photon index from the bright pulsar PSR~J1420-6048.}
\label{fig:K3_index_change}
\end{figure}

In order to investigate the spatial dependence of the spectral shapes, we separated the extraction region into four annular regions, i.e. "ring1": 0--$0^{\prime}.5$, "ring2": $0^{\prime}.5$--$1^{\prime}.5$, "ring3": $1^{\prime}.5$--$2^{\prime}.5$, "ring4": $2^{\prime}.5$--$3^{\prime}.5$.
To reduce the uncertainty of the column density with limited statistics, we fixed the absorption column density at ${\rm N_H}=4.4\times10^{22}~{\rm cm}^{-2}$ determined from the "ring1+ring2" region. Fig. \ref{fig:K3_4specs} shows the spectra (XIS0+3) for each circular region. In order to exclude the thermal contamination, we used the 2-10 keV band for the spectral fitting. In the "ring3" spectrum, we can see a weak trend of the iron emission line around 6.7~keV. This might come from unresolved thermal sources.
The photon index vs. relative coordinate is shown in Fig. \ref{fig:K3_index_change}.
Quoted errors are at the $1\sigma$ confidence level. We can see that the photon index slightly changes according to the distance from the bright pulsar region. 
The contamination effect from a point source in "ring1" on the outer region "ring4" was estimated as $\sim$5\%.
The fitted parameters are summarized in Table \ref{tab:fit}.
As for the HXD-PIN, we could not detect any significant pulse in timing analysis, thus there is no excess 
in the hard X-ray band from the pulsar.

\subsubsection*{Rabbit}
\begin{figure}
 \includegraphics[width=0.9\hsize,clip]{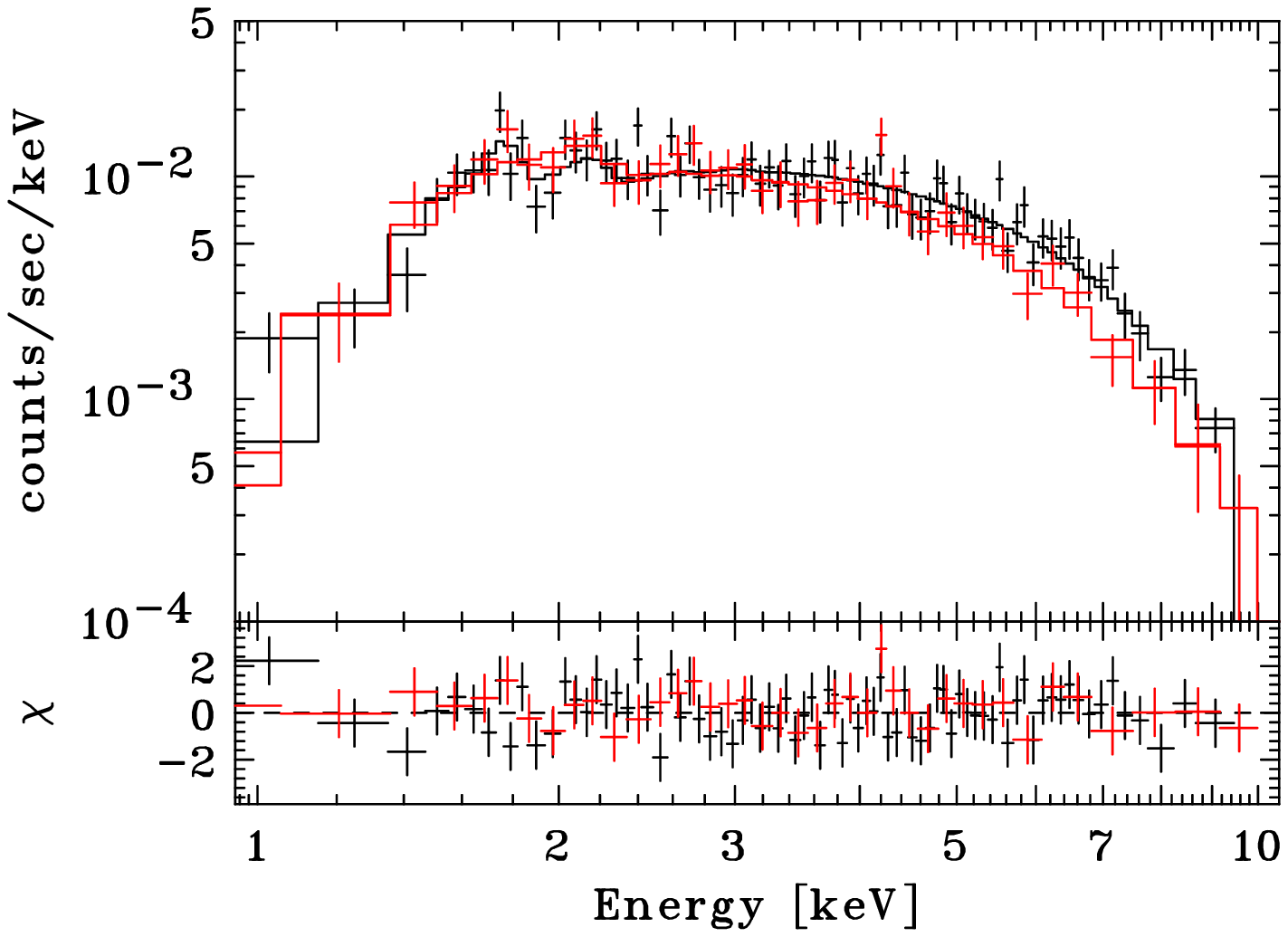}
\caption{Spectrum of the "ring1+ring2" in Rabbit (black: XIS0+3, red: XIS1). Solid lines show the fitted absorbed power-law model. The bottom panels show residuals from the best-fit.}
\label{fig:spec_Ra_src}
\end{figure}

We performed the same analytical procedures to the photon events from the Rabbit nebula as described in K3/PSR~J1420-6048.
Fig. \ref{fig:spec_Ra_src} shows the spectra extracted within an angular distance of $1^{\prime}.66$ from the X-ray peak, which corresponds to the "ring1+ring2" region in Fig. \ref{fig:Raimage} (b).
The background spectrum was consistent with that of the blank sky data.
The spectral shapes are well represented by an absorbed power-law model.
The absorption column density was determined as ${\rm N_H}=2.7\pm0.2\times10^{22}~{\rm cm}^{-2}$ by the spectral fitting.

\begin{figure}
 \includegraphics[width=0.9\hsize,clip]{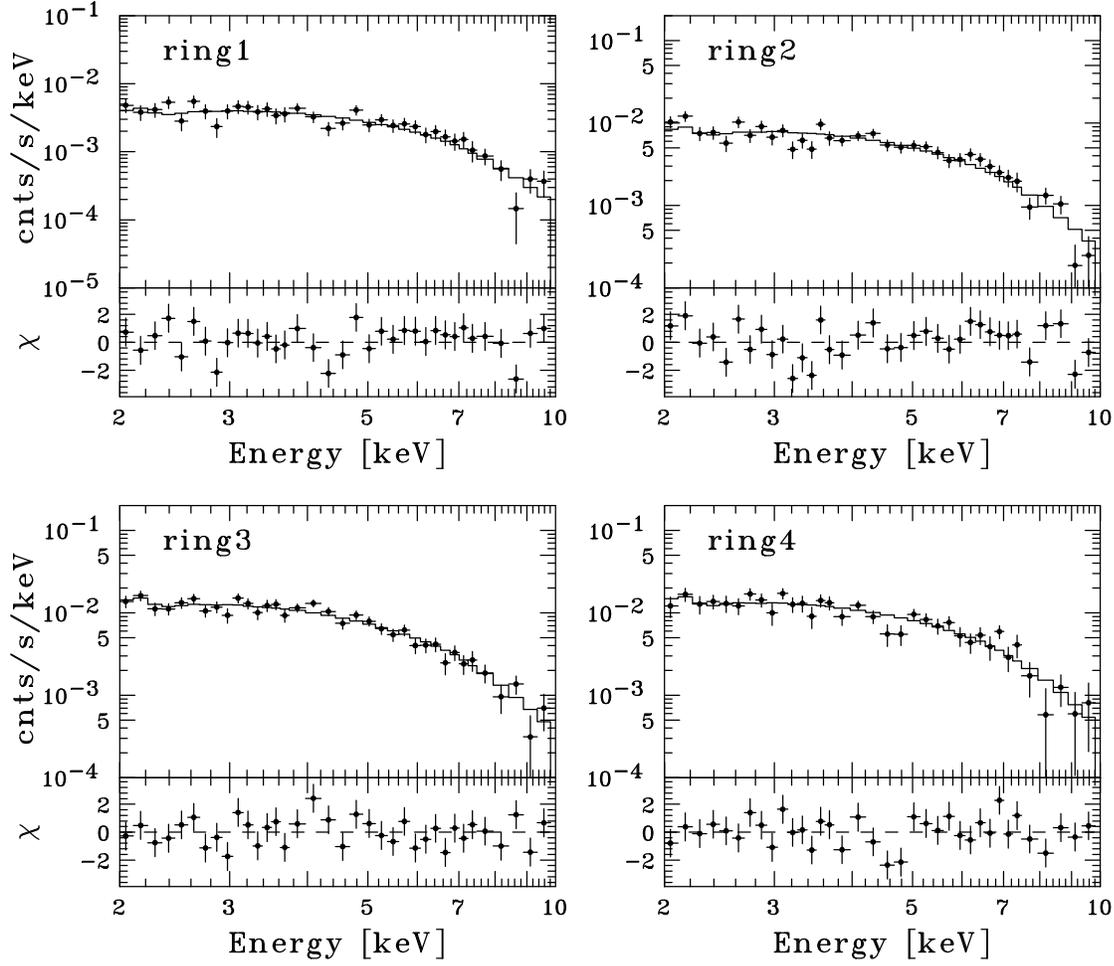}
\caption{Spectra of the circular regions in the Rabbit nebula. The extracted regions are shown in Fig. \ref{fig:Raimage} (b).}
\label{fig:Rabbit_spec_4circulars}
\end{figure}

\begin{figure}
 \includegraphics[width=0.9\hsize,clip]{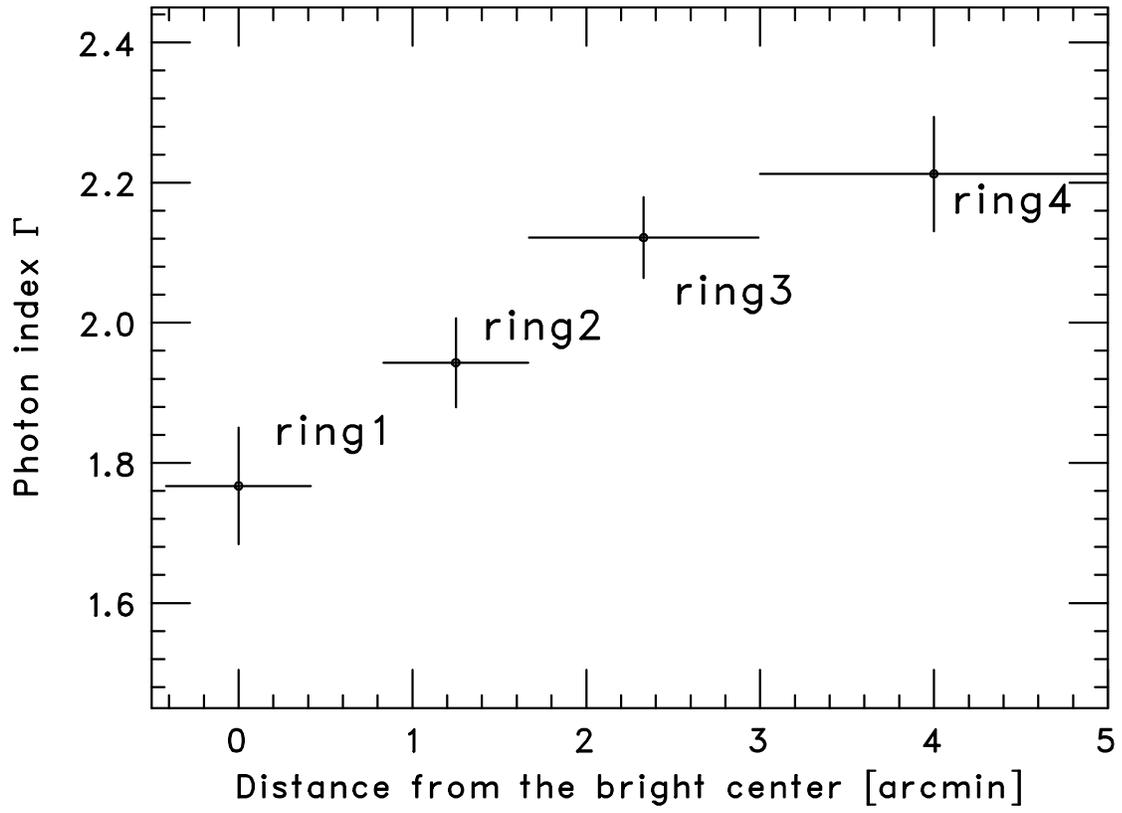}
\caption{Spatial dependence of the photon index from the bright center region in the Rabbit nebula.}
\label{fig:Rabbit_index_change}
\end{figure}

For the spatial dependence of the spectral shapes, we chose the annular regions as "ring1": 0--$0^{\prime}.83$, "ring2": $0^{\prime}.83$--$1^{\prime}.66$, "ring3": $1^{\prime}.66$--$3^{\prime}.0$, "ring4": $3^{\prime}.0$--$5^{\prime}.0$.
The spectral results are shown in Fig. \ref{fig:Rabbit_spec_4circulars}.
To exclude the thermal contamination, we used the energy range of 2--10~keV for spectral fittings.
The photon index vs. relative coordinate is shown in Fig. \ref{fig:Rabbit_index_change}.
The spectral shape becomes softer according to the distance from the inner bright core to the outer region, which is a similar trend as in the K3//PSR~J1420-6048 region. 
The systematic error of the PSF in "ring1" to "ring4" was less than 5\%.
The best-fit parameters are summarized in Table \ref{tab:fit}. The total flux in the energy range of 2--10~keV is estimated as $6.27\pm0.13\times10^{-12}$~erg$\cdot{\rm cm}^{-2}\cdot{\rm s}^{-1}$ determined from the extraction region of "reg1-reg2". 


\begin{table*}
\tabletypesize{\tiny}
\caption{Results of {\it Suzaku} XIS Spectral Fitting}
\begin{center}
\begin{tabular}{rrrrr}
\hline
\hline 
Object  \& region & $N_{\rm H}$~(wabs) \tablenotemark{a, b}  & Photon index \tablenotemark{a} & Flux [2-10~keV]  \tablenotemark{a,c}& $\chi^2_{\nu}(\nu)$ \\
 \hline
K3/PSR~J1420-6048, ring1+ring2 & 4.4 $\pm$ 0.3 & 2.00 $\pm$ 0.10 & 1.53 $\pm$ 0.03  & 1.01 (72) \\
reg1-reg2 & 4.4 (fix)& $2.17\pm0.06$ & $4.56\pm0.1$& 0.83 (33) \\\\ 
ring1 & ---& $1.69\pm0.16$ && 0.86 (32) \\
ring2 & ---& $2.06\pm0.06$ && 1.01 (32) \\
ring3 & ---&$2.28\pm0.07$ && 0.86 (35)  \\
ring4 & ---&$2.29\pm0.09$ && 0.94 (35) \\
\hline
Rabbit, ring1+ring2 &   2.7 $\pm$ 0.2 &  1.82 $\pm$ 0.10 & 2.65 $\pm$ 0.05  &  0.87 (101) \\
reg1-reg2 & 2.7 (fix) & 2.00 $\pm$ 0.06 & 6.27$\pm$ 0.13  &  0.80 (52) \\\\
ring1 & --- &$1.77\pm0.08$ && 1.06 (46) \\
ring2 & --- &$1.94\pm0.06$ && 1.09 (35) \\
ring3 & --- &$2.12\pm0.06$ && 1.10 (46) \\
ring4 & --- &$2.21\pm0.08$ && 1.08 (33) \\
 \hline
 \tablenotetext{a}{Quoted errors are at the 1$\sigma$ confidence level.}
\tablenotetext{b}{$N_H$ units are $10^{22}~{\rm cm}^{-2}$.}
\tablenotetext{c}{Unabsorbed fluxes are given in units of $10^{-12}$~erg~${\rm cm}^{-2}$~${\rm s}^{-1}$. }
\end{tabular}
\end{center}
\label{tab:fit}
\end{table*}

\section{DISCUSSION}
\subsection{Origin of the X-ray and $\gamma$-ray Emission}
We have detected the diffuse X-ray emission around K3/PSR~J1420-6048 and Rabbit, which were unnoticed in the previous observations, in the Kookaburra region with the {\it Suzaku} satellite. 
The X-ray peaks of the two sources are both within the error circles of the TeV sources and each of them has an offset from the $\gamma$-ray peak with $2^{\prime}.8^{+0.5}_{-0.6}$ for K3/PSR~J1420-6048 and $4^{\prime}.5^{+0.8}_{-0.9}$ for Rabbit. The XIS spectra of the two sources were well reproduced by an absorbed power-law model with a photon index of $\Gamma=1.7$ -- 2.3.

If TeV/X emission comes from the same object, the origin of the VHE $\gamma$-rays can be explained by the inverse Compton (IC) scattering of the CMB photons by high-energy electrons, while the X-ray emission via synchrotron radiation
in a mean magnetic field $B$. 
In this case, typical energies of responsible electrons are 
$E_e \simeq 20\cdot \bigl( \frac{\epsilon_{{\rm IC}}}{1~{\rm TeV} } \bigr)^{1/2}$~TeV for IC $\gamma$-rays (in the Thomson regime) at a photon energy $\epsilon_{\rm IC}$, 
and 
$E_e \simeq 70\cdot \bigl(\frac{B}{10~\mu G}\bigr)^{-1/2}\cdot \bigl(\frac{\epsilon_{\rm syn}}{1~\rm keV}\bigr)^{1/2}$ TeV for X-rays at $\epsilon_{\rm syn}$, 
 respectively. 
The synchrotron and IC spectra produced by the relativistic electrons 
obeying an energy distribution of $N_{e}(E) \propto E^{-p}$, have  the 
same spectral shape, $N_{\rm sync~or~IC}(\epsilon_{\rm syn~or~IC}) \propto \epsilon_{\rm syn~or~IC}^{-\frac{p+1}{2}}$ as long as IC scattering occurs in the Thomson regime. 
If the high-energy electrons are continuously injected into the radiation zone at a constant rate and loosing energy predominantly by 
synchrotron or IC (Thomson) losses, 
the electron energy spectrum becomes one power of $E$ steeper 
above the cooling break $E_{\rm br}$ that is determined by the cooling timescale 
and the age of the source.
 As a result, the synchrotron and IC spectra from the cooled electrons become softer by $\Delta\Gamma=0.5$. 

Our {\it Suzaku} results clearly show that the photon index smoothly changes according to the distance from the bright center regions, and the X-ray photon indices near the $\gamma$-ray peaks are roughly consistent with those in the TeV energies in both objects ($\Gamma_{\rm TeV} \simeq 2.17\pm0.1$ for HESS~J1420-607 and $\Gamma_{\rm TeV}\simeq 2.22\pm0.1$ for HESS~J1418-609, \cite{aha06}).
In addition to that, the difference of the photon index between the bright center and diffuse emission is $\Delta\Gamma \simeq 0.5$. 
The smooth spectral steepening of the synchrotron X-ray emission 
likely reflects  synchrotron burn-off of  the accelerated electrons.
Such effects have been observed for some young PWNe \citep{gaensler}, 
but we now find an interesting example of the smoothly steepening of 
the synchrotron X-ray emission in a middle-aged PWN 
of K3/PSR~J1420$-$6048 ($\tau_{\rm sd} \sim 13$ kyr).

In order to estimate the mean magnetic field strength, we 
modeled  the spectral energy distribution.
Since information about the spatial dependence of 
 the TeV emission is not available, 
 we assumed a simple one-zone synchrotron + IC model as the first order approximation, in which a single population of relativistic electrons in TeV energy range emit both X-rays through synchrotron radiation and TeV $\gamma$-rays through IC scattering off the CMB. 
The electrons energy distribution is formally assumed to 
be an exponentially cutoff power-law of the form 
$N_e \propto \gamma^{-p} \exp (-\gamma/\gamma_{\rm m})$, 
where $\gamma = E_e/m_e c^2$. 
We do not specify the physical meaning of $\gamma_{\rm m}$, which may 
account for both the maximum electron energy and the cooling break. 
This spectral form is invoked only for the purpose of an estimate of the 
magnetic field.

\begin{figure*}
\begin{center}
 \includegraphics[width=0.9\hsize,clip]{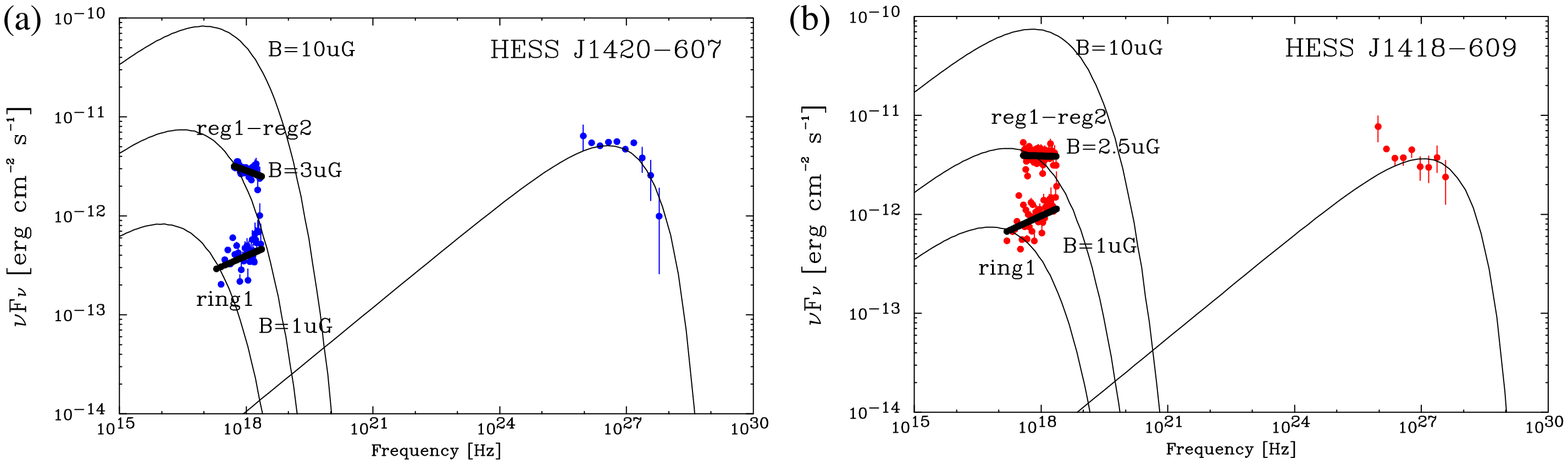}
\caption{Spectral energy distribution for the {\it Suzaku} and H.E.S.S. measurements with the synchrotron and inverse Compton models. Black thick lines indicate the best-fit results with absorbed power-law models. Electron spectral index is assumed to be $p$=2.3. The normalization and the maximum energy of electrons are adjusted to reproduce approximately the X-ray data and the TeV data. Black thin lines show the different ambient magnetic fields.}
\label{fig:sed}
\end{center}
\end{figure*}

{\bf Fig. \ref{fig:sed} shows the spectral energy distribution of HESS~J1420-607 and HESS~J1418-609 at the X-ray and $\gamma$-ray bands; the model curves 
are obtained with a combination of $\gamma_{\rm m} =8\times10^{7}$ (HESS~J1420-607) or $\gamma_{\rm m} =2\times10^{8}$ (HESS~J1418-609) and $p=2.3$ for both objects. 
The black thick lines overlaid on the {\it Suzaku} data points indicate the absorbed power-law models with best-fit parameters as shown in Table \ref{tab:fit}. 
We note that the "ring1" region mainly comes from bright pulsars while the "reg1-reg2" region includes both emissions from pulsar and diffuse components. 
The deviation between the spectrum of the "reg1-reg2" region and the model curve in higher X-ray energies comes from the pulsar's intrinsic hard spectrum since the intrinsic diffuse emission can be approximately estimated as "reg1-reg2" - "ring1".}
The magnetic fields can be estimated as $3~\mu$G for HESS~J1420-607 and $2.5~\mu$G for HESS~J1418-609. 
These values are in agreement with the magnetic field strength
that is expected for middle-aged PWNe, 
which lends support to the PWN scenarios for both objects.


\subsection{Morphological properties}
The diffuse X-ray emission of K3/PSR~J1420-6048 
has asymmetry (a circular shape of the radius $R_{\rm diffuse}(3\sigma_{\rm X})=8.1 \pm 0.4$~pc + a tail-like component) and its tail-like component extends to the $\gamma$-ray peak. On the other hand, the diffuse emission in the Rabbit seems to be more concentrated near the bright central source with higher surface brightness and relatively symmetrical shape of $R_{\rm diffuse}(3\sigma_{\rm X})=6.5 \pm 0.3$~pc. 
The angular sizes of the VHE $\gamma$-ray sources 
are larger than those of the X-ray emission regions, e.g. $R_{\rm diffuse}(3\sigma_{\rm TeV})=16.1\pm2.4$~pc (HESS~J1420-607) and $R_{\rm diffuse}(3\sigma_{\rm TeV})=14.8\pm 0.9$~pc (HESS~J1418-609).


For the estimated magnetic field of $B\sim 3\mu$G,
 the energy loss rate due to synchrotron radiation is 
just comparable to the loss rate via 
IC scattering off the CMB in the Thomson regime, which is given by
\begin{equation}
\tau_{\rm cool} \sim 140 \Bigl(\frac{E_e}{10~{\rm TeV}}\Bigr)^{-1}~{\rm kyr}.
\end{equation}
The lifetime of the electrons emitting synchrotron X-rays at 1 keV is 
$\tau_{\rm syn}(\epsilon = 1\, {\rm keV}) \sim 10$ kyr for the estimated 
magnetic field strength.
On the other hand, the energy loss timescale of the electrons emitting 
IC $\gamma$-rays at 1 TeV is 
$\tau_{\rm IC}(\epsilon = 1\, {\rm TeV}) \sim 70$ kyr \citep[see e.g.,][]{jager}. 
The TeV emitting electrons have a longer lifetime than 
the cooling time of the X-ray emitting electrons, 
which is shorter than the characteristic age of PSR~J1420$-$6048 
and probably also shorter than that of the pulsar in the Rabbit.
Therefore the morphological differences between the two bands 
can be expected, though details may depend largely on 
 the PWN evolution \citep{bambab, vink, etten}.

According to \cite{mattana}, $\gamma$-ray to X-ray energy flux ratio of PWNe is proportional to the characteristic age. The energy flux ratios for both objects are
$F_{\gamma}(1-30~{\rm TeV})/F_{X}(2-10~{\rm keV})\sim3.2$ for PSR~J1420-6048 and $F_{\gamma}(1-30~{\rm TeV})/F_{X}(2-10~{\rm keV})\sim0.16$ for the Rabbit nebula ($F_{\gamma}(1-30~{\rm TeV}$) values were taken from \cite{mattana}).
If we assume that the characteristic ages for both objects are correct, i.e. $\tau=13$~kyr for PSR~J1420-6048, and $\tau=1.6$~kyr 
for Rabbit, the flux ratios roughly follow the Mattana's relation. 


\subsection{Diffusion of Electrons}
The high-energy electrons are thought to be transported via convection and/or diffusion due to the magnetic fields randomized at the termination shock.
Let us assume that diffusion is the dominant transport mechanism. 
Indeed the pulsar wind slows down from relativistic to relatively low expansion velocities at some distance after the pulsar wind has been shocked.
The diffusion coefficient can be estimated as 
\begin{equation}
\kappa \sim R_{\rm diffuse}^2/\tau_{\rm syn},
\end{equation}
 where $R_{\rm diffuse}$ is the observed size of the synchrotron X-ray nebula. 
On the other hand, the diffusion coefficient can be written as 
\begin{eqnarray}
\kappa & \sim & \lambda c/3 \\
            & \sim & 2.6\times 10^{26} \cdot f \cdot 
\left(\frac{B}{10~\mu G}\right)^{-1}\left(\frac{E_e}{70~{\rm TeV}}\right)\ 
{\rm cm}^2\, {\rm s}^{-1}. 
\end{eqnarray}
where $\lambda = fr_{\rm L}$ is the mean free path of scattering, 
parameterized by a gyrofactor $f$ and the electron gyroradius 
$r_{\rm L}=E_{e}/(eB)$. 
By equating 
Eq.~(2) with Eq.~(3), {\bf we can estimate the gyrofactor as $f\sim1$ for both objects, which indicates that the quite efficient acceleration has occurred in these PWNe.}
 Interestingly, diffusion of electrons is slow, 
 nearly at the  the Bohm limit of $f=1$.
 The presumed configuration of 
the toroidal magnetic field in PWNe means that 
the direction of the magnetic field in the post-shock flow should be 
essentially perpendicular to the radial flow direction.
Therefore, TeV electrons can be confined quite efficiently in the absence 
of a radial field component. In this regard, the slow diffusion 
is not unexpected for PWNe.
 If instead diffusion of electrons occurs along 
a radial magnetic field line, the magnetic field in the post-shock flow 
must be highly turbulent so that the Bohm limit is realized. 

To further investigate diffusion processes, we need to incorporate 
spatial and temporal evolution of the physical parameters of the 
post-shock flow into morphological and spectral modeling. 
We expect that the next-generation Cherenkov telescopes, 
Cherenkov Telescope Array (CTA), will make it possible to 
perform spatially-resolved spectroscopy at the TeV $\gamma$-ray band, 
allowing us to constrain the spatial and temporal evolution of the PWNe 
by combining X-ray and TeV data \citep[see][]{etten}.



%

%
\acknowledgments
We thank an anonymous referee for helpful comments. Y. T. is supported by research fellowships of the Japan Society for the Promotion of Science for Young Scientists.

\end{document}